\documentclass[journal,onecolumn]{IEEEtran}
\usepackage[T1]{fontenc}
\usepackage[latin9]{inputenc}
\usepackage{color}
\usepackage{float}
\usepackage{amsmath}
\usepackage{amssymb}
\usepackage{graphicx}
\usepackage{csquotes}

\makeatletter

\floatstyle{ruled}
\newfloat{algorithm}{tbp}{loa}
\providecommand{\algorithmname}{Algorithm}
\floatname{algorithm}{\protect\algorithmname}

 \usepackage{algolyx}

\@ifundefined{showcaptionsetup}{}{%
 \PassOptionsToPackage{caption=false}{subfig}}
\usepackage{subfig}
\makeatother

\begin{document}

\makeatletter
\newcommand{\specificthanks}[1]{\@fnsymbol{#1}}
\makeatother

\title{Towards Bi-Directional Communication in Human-Swarm Teaming: A Survey}

\author{Aya Hussein, Leo Ghignone, Tung Nguyen, Nima Salimi, Hung Nguyen, Min Wang, and~Hussein A. Abbass,~\IEEEmembership{Senior Member,~IEEE,}
\thanks{The authors are with the Trusted Autonomy Group, School of Engineering and Information Technology, University of New South Wales, Canberra, ACT 2600, Australia.}
}

\markboth{}%
{Goh \MakeLowercase{\textit{et al.}}: Bare Demo of IEEEtran.cls for Journals}

\maketitle
\begin{abstract}

Swarm systems consist of large numbers of robots that collaborate autonomously. With an appropriate level of human control, swarm systems could be applied in a variety of contexts ranging from search-and-rescue situations to Cyber defence. The two decision making cycles of swarms and humans operate on two different time-scales, where the former is normally orders of magnitude faster than the latter. Closing the loop at the intersection of these two cycles will create fast and adaptive human-swarm teaming networks. This paper brings desperate pieces of the ground work in this research area together to review this multidisciplinary literature. We conclude with a framework to synthesize the findings and summarize the multi-modal indicators needed for closed-loop human-swarm adaptive systems.

\end{abstract}



\IEEEpeerreviewmaketitle{}

\section{Introduction}

\emph{Human-machine teaming} (HMT) involves concurrent interactions among humans and machines. A swarm is a group of distributed machines able to self-organize and generate group-level emergent behaviors from decentralized local interactions. \emph{Human-swarm teaming} (HST) extends HMT to interactions with a swarm. We will reserve the concept of \emph{Human-swarm interaction} (HSI) to situations where emphasis is required for the interaction dimension, while HMT will be used when emphasis is required for the \emph{teaming} dimension, and HST where the context necessitates both swarm and teaming interactions.

HMT operates within a context defined by a \emph{mission}. For clarity, a mission is defined by a set of objectives to be pursued by both humans and machines. Each of these objectives can be optimized through the completion of a number of tasks, which can be recursively divided into sub-tasks, and potentially could be adaptively determined while pursuing mission's objectives, while allowing for the team to negotiate plans adaptively and through their interactions. Within the context of this paper, we will assume that the overall mission has fixed and definitive objectives. Examples of these fixed objectives is to save the maximum number of lives in a search and rescue scenario or to maximize the area coverage by a swarm of air vehicles surveying a mine site. These objectives are on the HMT level.

Dyer \cite{dyer1984team} offers the basic ontological constituents of a `team' as \enquote{social members}, \enquote{task interdependency}, and \enquote{shared goals}. Subsequent literature \cite{morgan1993analysis,katzenbach1993discipline} added \enquote{adaptive interaction} and \enquote{commitment} from different members in the team with distinctive skills towards performance improvement and accountability. Teaming relies on teamwork skills such as clarifying interdependencies, establishing trust, and finding out means for coordination \cite{edmondson2012teaming}. Castellan \cite{castellan2013individual} discussed the functional requirements for team members as clearly defined roles and responsibilities, task-related knowledge, and interdependent connection between one another. These three dimensions of a team can be used to distinguish teams from swarms in which members are homogeneous regarding expertise, roles, and responsibilities. The dynamic concept of \emph{teaming} involves coordination and collaboration activities with flexible team structures.

Bringing together the concept of \enquote{teaming} in HST raises a number of scientific challenges. Some of the challenges rest on the design of appropriate artificial intelligence algorithms to allow the swarm to be smart enough to collaborate with the human. Some challenges are epistemological in nature and call for a better understanding of distributed cognition and the form of distributed situation awareness within a swarm. 

This paper focuses, however, on a third form of challenges; the bidirectional communication that needs to take place between the humans and the swarms and the artificial intelligence agents that need to adapt and orchestrate this interaction. This third challenge sits at the core of the first two. Without bidirectional communication, the human-swarm teams will fall short in their abilities to effectively and efficiently collaborate. Without smart agents for bi-directional communication, the swarm and humans will find it difficult to collaborate and/or coordinate actions. Even in the simplest HST systems, a basic form of these agents are needed and may take different forms, from a pre-programmed graphical user interface to human-friendly natural communication mediums such as voice and gesture.

In section~\ref{sec-HST} we discuss the concept of HMT and its properties, then we review possible autonomy configurations for HST, discussing risks and remedies. In section III, we distill five groups of indicators that are needed at the human-swarm interface, namely mission performance, interaction, mission complexity, swarm automation, and human states indicators. These indicators are complementary to each other and can reflect the states of human, swarm, interaction, and mission effectiveness. In section IV we synthesize together these five groups of indicators into the MICAH framework, representing a synthesis of the literature to enable flexible and intelligent adaptive control. The closed-loop adaptive system based on MICAH for human-swarm teaming is then discussed. The paper concludes with the challenges and open research questions towards achieving effective HST in section V.

\section{Human-Swarm Teaming}\label{sec-HST}

The field of human-robot interaction (HRI) has offered a wide range of approaches towards effective and efficient control interfaces among a group of multiple agents. Extending HRI to HMT, decisions get jointly made by both the human operator(s) and machines. When the teaming relationship evolves to shared control tasks involving mixed-initiatives, it becomes pertinent to develop natural and seamless bi-directional communication between the humans and the machines. Such a requirement comes with a few challenges including how to dynamically adjust the level of autonomy of different players, how to strengthen mutual trust, and what mechanisms are required to facilitate situational awareness (SA) of team members~\cite{chen2014human}.

In this paper, we investigate the teaming of human operators with swarms of multiple agents, or HST for short. In particular, the paper will focus on HST contexts where humans and swarms need to work together smartly and communicate sufficient information to achieve a stable bi-directional communication for shared situation awareness, shared mental models, team mutual predictability, and adaptability. At the interface of this bi-directional communication sits a smart agent that needs to manage the states of different entities in the system. We will first start with the criteria needed to create and effective team.

\subsection{Human-Machine Teaming: Criteria for Effective Teaming}

We can borrow some of the key factors for effective HMT from social science research conducted on effective human teams~\cite{salas2008teams}. Examples of these factors include shared cognition, team training, and collective orientation. \emph{Shared cognition} includes shared mental models, team situation awareness, and comprehensive communication. \emph{Team training} promotes teamwork and enhances team performance. Developing a \emph{collective orientation} unifies team coherence and intent.

Similar factors were found in HMT studies including \emph{Mutual predictability}, \emph{shared understanding}, and \emph{adaptability} to other team members~\cite{sycara2004integrating}. These key properties of a HMT system facilitate the coordination and cooperation of machines with human teammates. Both parties have to recognize their teammate's actions to infer their intentions corresponding to the context. To develop such a capability on the machine side, it requires sociocognitive mechanisms that support the interpretation of mental processes of the partners during interactions to improve the effectiveness and efficiency of dynamic planning and behavioral adaptation to achieve shared goals~\cite{wiltshire2013towards}.

Sycara and Sukthankar \cite{sycara2006literature} identified that information exchange, communication, supporting behaviour, and team leadership play vital roles in effective HMT. More recent literature expanded the list of factors for an effective HMT to seven: Belief in concept of a team, communication, leadership, performance monitoring and feedback, coordination and interdependence, situational awareness, awareness of individual's unique cognitive model and shared cognitive model \cite{joe2014identifying}.

\subsection{Autonomy in Human-Swarm Teaming}

In HST, the human and the swarm are assigned complementary roles with the aim of combining their skills efficiently and in a manner that achieves mission success and an efficient overall team performance. Fixing the level of autonomy within the swarm produces a rigid system with little smartness. Such setting can overload the human as the human needs to fill the gap between task requirements and the swarm fixed design. This applies regardless of the level of autonomy the swarm exhibits. If the level of autonomy is low, the human carries most of the load with the end result finding the human overloaded. If the level of swarm autonomy is high, the human could become under-loaded. Both situations are undesirable and can lead to difficulties in sustaining human situational awareness (SA) and a drop in human and system-level performance  \cite{endsley1997} and engagement level. Fixed autonomy has been criticized for its negative consequences on cognitive load, SA, and performance. Furthermore, fixed autonomy has been associated with human's complacency and skill degradation \cite{aiding}. For example, in a complacency state, human might not be able to detect automation failures in multi-tasking environments \cite{parasuraman1993performance}.

Flexible autonomy begs for a degree of smartness. Chen et al. \cite{teaming} classified flexible autonomy approaches into three classes: adjustable, adaptive, and mixed-initiative, based on whether the adaptation decision is taken by the human, the software, or both. 

In adjustable autonomy the human initiates the joint human-system task adaptivity \cite{chen2014human}, \cite{miller2007designing}. One of the disadvantages associated with adjustable autonomy is that task delegation by human could be very time consuming and potentially unsafe in situations where the human operator is naturally overload because of task demands \cite{chen2014human}. Although adjustable autonomy could be improved using feedback through the human-machine interface, belated decisions made by a human could decrease the overall performance of the mission. 

Task delegation in a human's hands is also subjective and depends on human factors that reach beyond the realm of workload (e.g. emotional stress). Some operators would place more constraints on the automation at the expense of time, while others might place less constraints on automation at the expense that the automation's behavior can diverge from the operator's intent \cite{miller2005playbook}. In addition, the management of task allocation by humans is vulnerable to automation-induced complacency. It is worth noting that the human operator is less likely to detect automation failure in constant-reliability conditions \cite{parasuraman1993performance}. High workload and fatigue could also contribute to the automation-induced complacency. If the type of task delegation is adjustable in human-swarm teaming, and the swarm has an initial constant successful performance, the human is unlikely to detect the failure of the swarm afterwards due to the complacency state.

In adaptive autonomy, the software decides on whether the level of autonomy should be risen or lowered based on some input indicator(s). The accuracy of this decision is important to avoid sudden changes that can be inappropriate or annoying \cite{teaming}. Results from different studies indicate that many factors can determine the appropriate level of autonomy. Abbass et al. \cite{AdaptiveHA} used task complexity and human workload to adapt the level of autonomy in air traffic control tasks. Rusnock et al. \cite{threshold} studied the use of different workload thresholds in adaptive autonomy. They found that the proper threshold depends on both the human and the task such that in some tasks, increasing this threshold results in an increase in both workload and SA, while in others, increasing the threshold results in an increase in workload without benefiting SA. Feigh et al. \cite{triggers} proposed a taxonomy of triggers that can be used for adaptation. They categorized the triggers into five categories: operator, system, environment, task/mission, and spatio-temporal triggers.

The third class of flexible autonomy is mixed-initiative systems in which the adaptation decision is taken collaboratively including input from both the human and the software agent. These systems combine the ideas of both adaptable and adaptive autonomy as they allow the human to intervene with the adaptation decision taken by the agent \cite{teaming}.

To generalize the review of this paper, we will focus on mix-initiatives, where both humans and machines control the level of autonomy; thus, both need to be smart enough to recognize their mutual states to make timely actions on when and how to adjust the level of autonomy.

Our work focuses on triggers that can be used for adaptation either in adaptive autonomy or mixed-initiative systems. This is similar to Feigh et al. \cite{triggers}, in that both works provide taxonomy of triggers. However, the difference between this paper and Feigh et al. one is fourfold. First, this paper focuses on how each of the adaptation triggers can be quantified using synthesized indicators from the literature. Second, our main interest is HST; thus, this paper extends Feigh et al. work using a swarm-lens. Third, we discuss how HSI indicators can be included in the adaptation decision to ensure that the selected autonomy configuration results in effective teaming between the human and the swarm, i.e. to ensure that human interventions are constructive and of added value to the performance. Finally, mission-level performance indicators are included with the aim of ensuring that the benefits of the autonomy configuration translate to mission-level improvements.

\section{Indicators for Adaptive HST Systems}
\subsection{Mission Performance}

Automation equips an automaton with functions to process and/or execute tasks. The level of automation, therefore, represents an agent's capacity to perform a task, while autonomy expresses \enquote{the \emph{freedom} to make decisions} \cite{abbass2016trusted} afforded by the opportunity that exists to allow an agent to act. Autonomy carries negative risks when the capacity of an agent, that is, automation, is conceptually less than the capacity required to perform a task given an opportunity within a mission. HMT brings humans as biological autonomous systems with autonomous machines to work together to optimize mission objectives.

The primary aim of the team composed by human and swarm is to perform the mission successfully. It is therefore of extreme importance to allow the team to monitor progress towards the mission objective(s) in order to take corrective actions and/or adapt accordingly. In this section, we review the literature from the lens of indicators for mission success. We distinguish between how to measure effectiveness (achieving mission success) and efficiency (achieving the success using minimum resources/time) of the system performance in human-swarm teaming. A comprehensive set of metrics can only be defined in terms of the specific tasks composing the mission at hand. Therefore, we will review some examples of mission types from the literature and the metrics that have been defined to evaluate them.

\subsubsection{Mission Performance in HRI}
Jacoff et al. \cite{jacoff2001reference} \cite{jacoff2001standard} under the umbrella of the National Institute of Standard and Technology (NIST) proposed a list of quantitative and qualitative metrics to evaluate the performance of autonomous ground vehicles in a search-rescue mission (e.g. number of victims localized, number of obstacles found, and recovery rate). 

Howard et al.~\cite{howard2002mobile} proposed a deployment approach to achieve broad area coverage. They experimented with 100 robots which get repelled by other robots and obstacles, thus spreading through the entire environment. A different approach is given Ganguli~\cite{ganguli2007visibility} where each robot can sense distances to the environmental boundary and other robots, and then get deployed to cover the entire environment. The coverage area is a metric that is used in these situations to measure mission success.

Olsen and Goodrich~\cite{olsen2003metrics} suggest to differentiate between overall mission effectiveness and current task effectiveness. Crandall and Cummings~\cite{crandall2007identifying} evaluate mission performance using two factors: the number of objects collected and the number of robots remaining. Efficiency is measured in terms of time using two questions: how much time it takes to achieve the mission? and the average time to complete all subtasks.

The operator is almost absent in the metrics discussed above. To overcome this problem, Scholtz et.al.~\cite{scholtz2004evaluation} designed the below operator-centric indicators and called them critical incidents: 
\begin{itemize}
\item Global Navigation: The operator needs to have a view about the area robots are working;
\item Local Navigation: The operator needs to know the environmental factors close to the robots in order to avoid some mistakes during interaction with robots hazards like doorways or trees;
\item Obstacle Encounter: The robots need to deal with obstacles while moving to the goal;
\item Victim Identification: The operator has to identify a victim. In some cases, because of the inaccurate sensor data, this may lead to the operator misidentifying a victim; and 
\item Vehicle State: Informing the device status (e.g. battery, broken sensors) of robots to the operator. If this information is correctly provided, the operator may be able to pass these challenges to achieve the mission.
\end{itemize}

Based on these critical incidents, Steinfeld et al.~\cite{steinfeld2006common} define five key measures of effectiveness:
\begin{itemize}
\item Percentage of navigation tasks successfully completed
\item Coverage of area
\item Deviation from planned route
\item Obstacles that were successfully avoided
\item Obstacles that were not avoided, but could be overcome
\end{itemize}

\subsubsection{Mission Performance in HSI}
HRI metrics could be transferred to human-swarm teaming. Nevertheless, HST comes with distinct properties as discussed by Hayes and Adams~\cite{hayes2014human}.  

Nunnally et al.~\cite{nunnally2012human} show that mission effectiveness and efficiency metrics degrade when information on swarm is lacking. Similar to HRI, some factors are individuated such as the mission completion percentage, or the completion time. Harriott et al.~\cite{harriott2014biologically} suggest the use of resource depletion as an objective measure. Here, resource depletion is used to quantify the irreversible consumption of limited resources by members of the swarm. Manning et al.~\cite{manning2015heuristic} also rely on resource depletion as a metric and extend the concept with \emph{timing} to capture mutual delay time, which affects the behaviour of the entire swarm. Two other metrics discussed in the research are \emph{Micro-level Movements} and \emph{Macro-Level Movements}. Level of overlap in neighborhoods is used for the first measure, and the elongation, which represents the rectangular structure of the swarm, is used for the second.

\begin{figure*}[pt]
\centering
\includegraphics[width=0.8\textwidth]{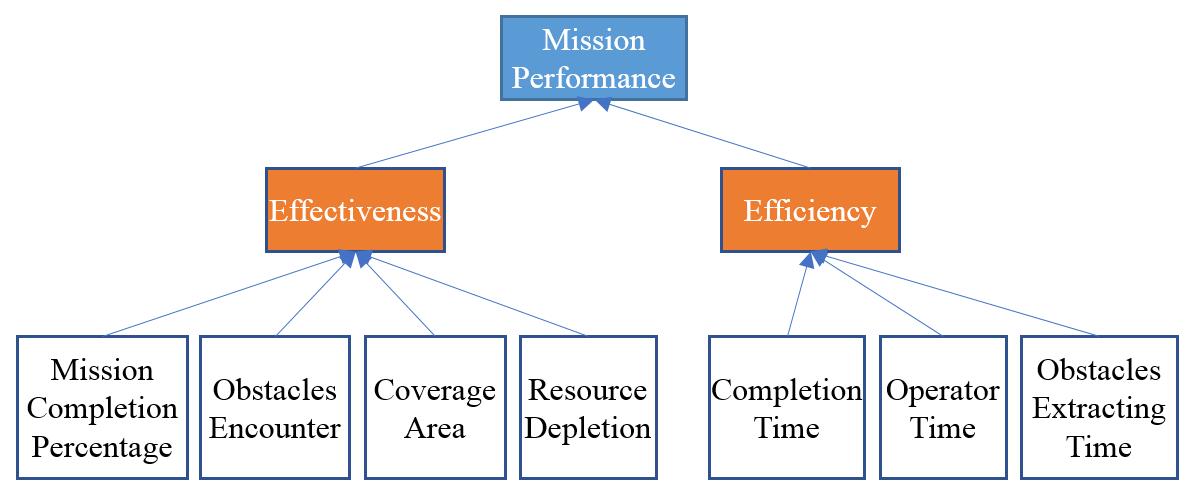}
\caption{Examples of the indicators useful for the evaluation of Mission Performance in different mission conditions, and their relations showed as a tree-graph.}
\label{fig:tree_mission_objs}
\end{figure*}

The indicators discussed above are distilled to form the tree presented in Fig~\ref{fig:tree_mission_objs}, where both measures of effectiveness and measures of efficiency form the two dimensions to measure \emph{Mission Performance}.

\subsection{Interaction Indicators}

The interaction between human and swarm refers to the communication approach and the control methods which allow for exchange of their intent information and actions. Factors that influence the interaction include level of autonomy, input timing, and neglect benevolence. A major challenge in HSI is the escalating complexity that could result from an increase swarm size and task demands. 

As the size of the swarm increases, the human has to monitor and control a larger group with massive number of interactions. 
For example, the human ability to control the swarm in a supervisor control task would severely be limited with 
the limited cognitive capacity of human operators~\cite{olsen2004fan}. Unfolding indicators for the effectiveness and efficiency of the interaction is important as both, a detection tool for when more or less automation is needed and as a diagnostic tool to understand the success or otherwise of the team. The remainder of this section will review interaction indicators.

\subsubsection{Basic Interaction Indicators}

In HST, it is necessary to identify the set of key metrics to represent the performance of the interaction, as well as the ability to predict the effectiveness of the interaction \cite{crandall2007identifying}. These key metrics can serve as the interaction indicators for an adaptive framework of HST.

\begin{figure}[tp]
\centering
\includegraphics[width=0.45\textwidth]{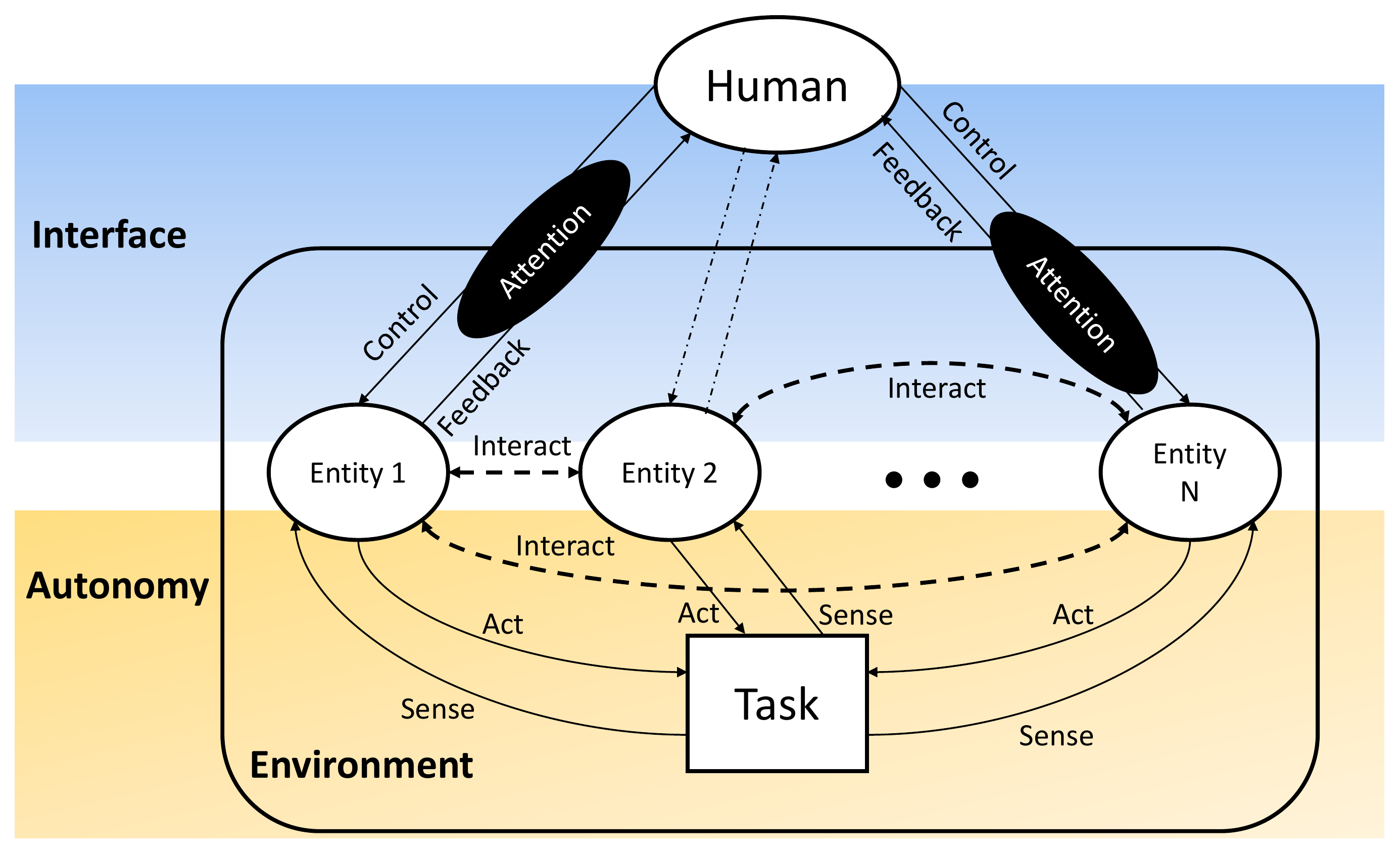}
    \caption{HSI including one single human and a swarm under supervisory-control.}
    \label{fig:interaction_loop}
\end{figure}

There are three fundamental metric classes used in HSI and were introduced in Crandall et al.\cite{crandall2007identifying}: \emph{interaction efficiency}, \emph{neglect efficiency}, and \emph{attention allocation efficiency}. Figure~\ref{fig:interaction_loop} shows the basic interaction loop for an HSI system. The efficiency of the interaction is commonly evaluated through interaction efficiency. The bottom loop describes the entities in the swarm. They sense the environment and produce appropriate actions corresponding to a certain level of autonomy. The efficiency of the entities performing the task without the attention of the human operator is assessed by neglect efficiency. The attention allocation efficiency is a metric class used to capture the efficiency with which the human operator allocates his/her attention among multiple entities. These three metric classes are dependent on one another and also dependent on the level of autonomy that the swarm component possesses.

\paragraph{Interaction Efficiency}

The interaction efficiency metric class comprises different metrics discussed in the literature. The most popular one is the \emph{interaction time} which is the amount of time needed for a human to manage one single entity in the swarm \cite{crandall2005validating}. When dealing with multiple entities in the environment, this metric can be extended to \cite{kerman2013methods}:
\begin{equation}
    \textit{IEm} = f(N(t))\times \textit{interaction\ \ time},
\end{equation}
where \textit{IEm} is the Interaction Efficiency for multiple entities and $N(t)$ is the number of agents the human interacts with at time $t$. The $f(N(t))$ term denotes a function describing the relationship between the swarm size and the time needed to manage the swarm. In the simplest, case this relationship might be linear in the increase of the number of agents in the swarm: $f(N(t)) = N(t)$.

\paragraph{Neglect Efficiency}
Basically the neglect efficiency can be assessed by the \emph{neglect tolerance} expressed by the time an agent can be ignored before the error exceeds a threshold~\cite{goodrich2003seven}. The neglect time has a direct relationship to the preservation of acceptable performance. Improving the neglect time is one goal of a successful HRI system, whereby the agent has enough capability to deal with the task. The neglect tolerance is not exactly an indicator that we use under the class of interaction indicators, but more under the class of automation indicators. However, we still mention this metric here because it has an indirect impact on reducing the \emph{interaction effort}.

\paragraph{Interaction Effort}
This metric can provide information of how a particular interface design affects the overall effectiveness of the interaction. Interaction effort is not only physically defined by the interaction time, but includes cognitive effort~\cite{olsen2003metrics} of subtask choice, information requirement of the new situation after a choice, planning, and intent translation. When interacting with multiple agents, the interaction effort can be estimated indirectly via neglect tolerance and Fan-out (the maximum number of agents the human is able to control effectively):
\begin{equation}
    IEft = \frac{neglect\ \ tolerance}{Fan-out \ \ -\ \ 1},
\end{equation}

\paragraph{Attention Allocation Efficiency}

When a human operates a swarm of multiple entities, the human must neglect some agents and focus his/her attention on controlling one agent. Therefore, the effectiveness of the HSI can also rely on another metric class called \emph{attention allocation efficiency}. This metric class includes SA of the entire swarm and environment, the switching time and the time the human makes decision on which agent to switch his/her attention to.

\subsubsection{Interventions}

Intervention metrics are used to estimate the cognitive and physical efforts of human when interacting with an autonomous robot in HRI. There are two kinds of interactions: planned and unplanned; unplanned interactions are understood as interventions~\cite{huang2003toward}.

Steinfeld et al.~\cite{steinfeld2006common} has referred to intervention metrics as \emph{non-planned interaction} metrics which can be used in robot navigation task. The intervention metrics include: the average number of interventions over a time period, the time required for interventions, and the effectiveness of intervention~\cite{scholtz2003evaluation}. The efficiency of the interaction can be also evaluated through the ratio of intervention time to autonomy time~\cite{yanco2004beyond}. For example, if the operator time is 1 minute to give a navigation instruction to robots, and then the robots complete the navigation task in 10 minutes, the ratio is 1:10.

Again, in the case of HSI, this category of metrics has a strong connection to the level of autonomy that the swarm component possesses. In the case there is shared control initiative and the possibility of negotiation between human and automation, it is essential to identify extra measures such as the percentage of requests for assistance created by robot, the percentage of requests for assistance created by human, and the number of insignificant interventions of human operator~\cite{steinfeld2006common}.

\subsubsection{Communication}

One practical problem in HSI is the factors impacting the communication channels between the human and the swarm including latency and bandwidth, especially in the case of teleoperation or remote interaction with a large swarm. The problem of limited bandwidth was mentioned in~\cite{mclurkin2006speaking} in an attempt to design an effective interface for HSI, in which the centralized user interface is responsible for human command broadcasting as well as integrating the information of the whole swarm to visualize them for the human operator. Kolling et al.~\cite{kolling2016human} reported a series of HSI experiments with different bandwidths. The conclusion supported the claim that the higher bandwidth offered larger capacity for multiple robots' states acquisitions in a time step. An increase in latency caused degradation of interactions~\cite{steinfeld2006common,walker2012neglect}. The problems mentioned above affect the effectiveness and the efficiency of the HSI because they impact the asynchrony of interaction among swarm members and delays in the bidirectional interactions. A solution for these problems can be a predictive display using swarm dynamics and bandwidth information.

\begin{figure*}[tp]
\centering
\includegraphics[width=0.80\textwidth]{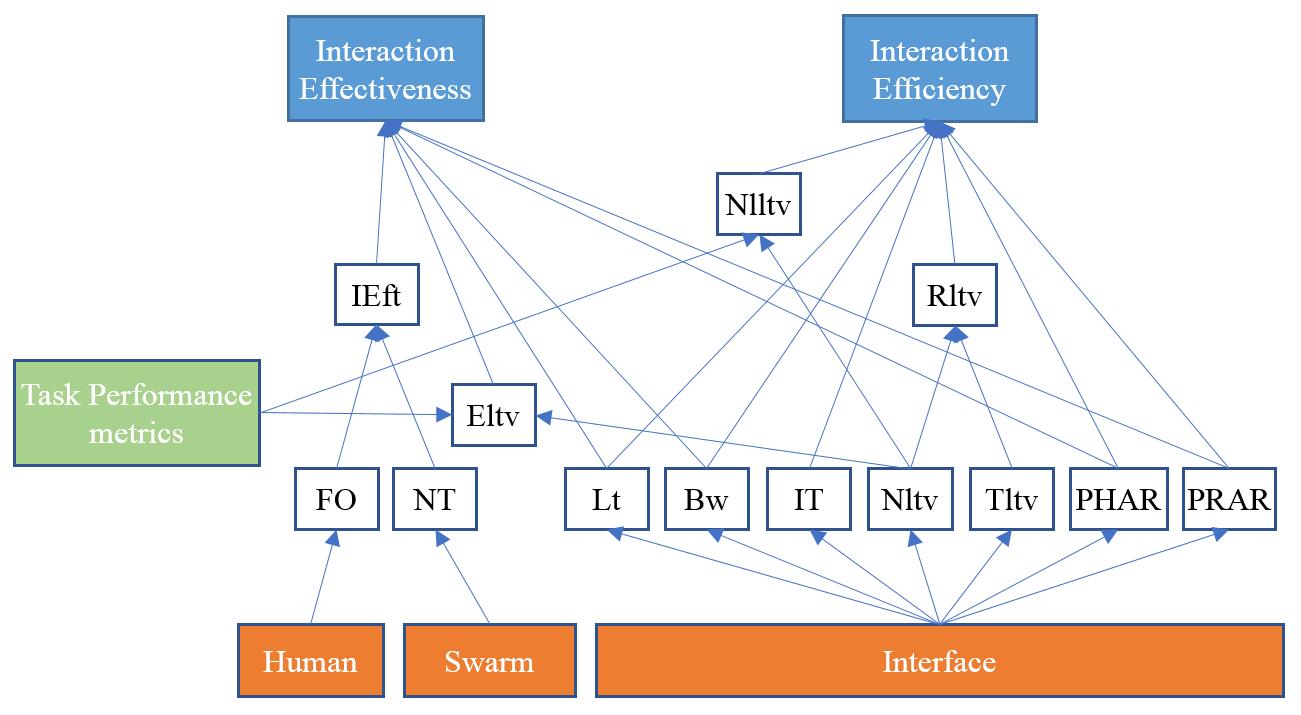}
\caption{A synthesized tree of interaction indicators for adaptive HSI system.}
\label{fig:tree_interaction}
\end{figure*}

The concepts introduced in this section are summarized in Figure \ref{fig:tree_interaction}, which presents the relationship between the top two interaction indicators, namely effectiveness and efficiency, with the sub-metrics and factors from human, automation and interface components.

\subsection{Mission Complexity}
Mission complexity could be defined as the amount of mental workload a mission will potentially exerts on a human.
Human mental workload can negatively hinder the success of a mission that relies on a collaboration between the human and the swarm. The persistent premise in related literature is that the nature of the mission impacts human mental workload. Both objective and subjective factors have an impact on the perceived mission complexity and the human performance~\cite{objectiveSubjective}. This section discusses only objective factors. We will divide the factors contributing to the complexity of a mission into three groups, depending on the component that generates these factors: either the swarm, the interface, or the structure of the mission.

Swarm characteristics can have a direct impact on human cognitive load. The level of autonomy of the swarm was shown to be an important factor of mission complexity. Ruff et al.~\cite{ruff2002} studied the workload associated with different levels of autonomy while controlling a group of four UAVs. They found that manual control resulted in the highest workload. Similarly, Riley et al.~\cite{usarJ} found that manual control of a robot resulted in a considerable workload in a search and rescue task. Mi et al.~\cite{LOA} also argued that manual operation dramatically increases the workload on the operator. However, semi-autonomous swarms also require devoting considerable cognitive resources as the human has to understand a plethora of information arriving from the swarm~\cite{swarmingNetwork} in order to maintain high level of situation awareness.

The size of the swarm can also result in increasing the workload requirements. Ruff et al. \cite{ruff2002} found that increasing the number of UAVs results in increasing the perceived workload. Furthermore, this increase is sharp in the case of manual control. However, by providing scalable control methods rather than controlling individual members, the workload can remain constant. Kolling et al. \cite{towards} proposed controlling the swarm in a foraging task using two control methods: selection and beacon. They showed that the number of human instructions didn't change significantly across different swarm sizes. Pendleton \cite{scalableHSI} used three control methods in a foraging task: leader, predator, or stakeholder. They found that using these control methods doesn't result in a significant change in the workload across different swarm sizes.

The interface between the human and the swarm can also contribute to mission difficulty. Pendleton \cite{scalableHSI} found that operator's workload is affected by the control method such that both control by a leader and a stakeholder results in a lower workload than control by a predator. 

Some research study the effect of information visualization on mission complexity. The amount of information presented can affect cognitive load such that too little information results in increasing uncertainty and leads humans to integrate information from other sources like their own assumptions, which could result in an increase in cognitive load~\cite{lost}. Too much information, on the other hand, caused information overload and makes the human overwhelmed with a large amount of information that may exceed their cognitive capacity. Van der Land et al.~\cite{SAChapter} argued that low-level information negatively impacts operators' cognitive load as they have to process it to build higher levels of SA~\cite{SAPerformance}. 

The method for presenting mission related information is equally important. This can be explained using cognitive fit theory which indicates that the efficiency of problem solving is enhanced  when there is a match between information presentation and the mission~\cite{cognitiveFit}, in which case the human uses the information directly without the need to mentally convert it into a representation that fits the mission. 

The choice of the display technology may have implications on  SA and workload. Ruiz et al. \cite{immersive} compared the use of different display technologies in multi-UAV operations. They found that virtual reality (VR) based immersive screen results in the best operator's SA and the lowest cognitive load. Besides, they found that while VR glasses outperform standard screen in terms of improving SA, this improvement comes at the cost of increasing workload.

Mission complexity can also result from the structure of the mission and how it is executed. For instance, the existence of sub-tasks that are executed concurrently adds to the human workload. Liu et al. \cite{complexity2012} pointed out that time constraints can result in task concurrency which leads to higher mission complexity by increasing the information load. Chen \cite{HATeaming} argued that switching between tasks can degrade performance as there can be interference between task related information. This interference increases if tasks are similar with respect to stimuli, processing stages, or required responses \cite{Caroline}. Moreover, the number of alerts or interruptions can have an impact on the cognitive load. Humans may need long time (up to 7 seconds) to recover from interruptions and switch back to the interrupted task \cite{HATeaming}.

\begin{figure*}[tp]
\centering
\includegraphics[width=0.80\textwidth]{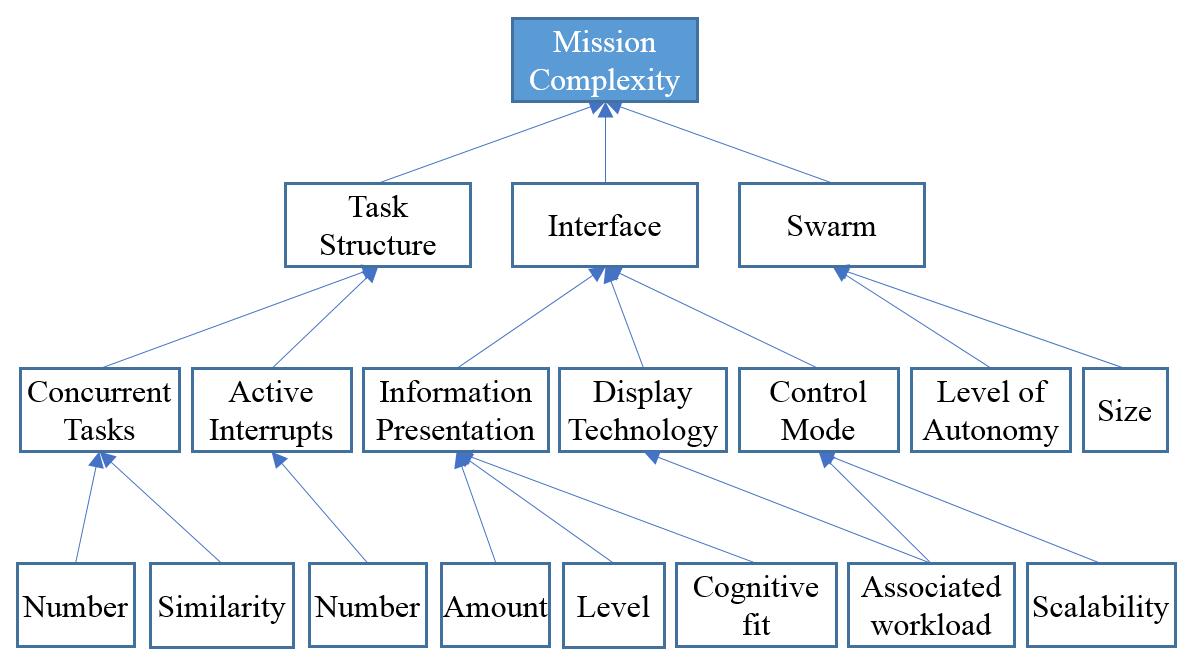}
\caption{A portfolio of Mission Complexity metrics.}
\label{fig:tree_complexity}
\end{figure*}

The concepts introduced in this section are composed in Fig~\ref{fig:tree_complexity}, showing how they can be combined to build an estimation of the \emph{Mission Complexity}.

\subsection{Swarm Automation Level}

Finding effective metrics for the analysis of a swarm, as stated in \cite{kolling2016human} is still an open problem. In order to define the degree of automation of a swarm we will start from the literature for Human-Robot Interaction, where automation is usually considered in terms of \emph{Neglect Tolerance} \cite{olsen2003metrics,crandall2007identifying}.

\emph{Neglect Tolerance} is often regarded as a static metric for the quality of an HRI system. Crandall et al.~\cite{crandall2005validating} presented a concept of \emph{Neglect Tolerance} which is slightly different from the one commonly found in literature, along with a method for evaluating it while the system is running. In order to preserve consistency both inside this paper and with the literature, we will report that idea with a slightly different terminology. For this purpose we introduce the concept of \emph{Human Dependence}, a measure of how much a robot is in immediate need of human intervention. This concept corresponds to the composition of two different sub-metrics: \emph{Neglect Tolerance}, which describes how the performance of the robot decreases while it is being neglected, and \emph{Interaction Efficiency}, which describes how the performance of the robot increases when a human starts interacting with it after a period of neglect. Both of these values are affected by the level of autonomy of the robot (e.g. a highly autonomous robot will not suffer much from being neglected, but will also experience reduced gains from human interaction), the complexity of the current situation, and previous history of interaction/neglect. The performance of a robot can then be described by the following equation:
\begin{equation}
  P(\pi,C,t)=
  \begin{cases}
    P_I(\pi,C,t_\textit{on},T_N), & \text{if interacting}\\
    P_N(\pi,C,t_\textit{off}),    & \text{otherwise}
  \end{cases}
\end{equation}
where $P$ denotes performance, $P_I$ denotes performance while the human is interacting with the robot, $P_N$ denotes  performance while the human is neglecting the robot, $\pi$ denotes current level of autonomy, $C$ denotes the complexity of the situation, $t_\textit{on}$ and $t_\textit{off}$ denote the times since the start of the current interaction/neglect, and $T_N$ denote the time the robot had been neglected before the start of the current interaction.

In the Human Swarm Interaction literature, it is possible to find metrics and techniques to estimate the complexity the swarm is currently dealing with. Some useful metrics are described in~\cite{manning2015heuristic} as follows:

\begin{itemize}
\item Cohesion: Evaluating the connectivity level of swarm.
\item Diffusion: Assessing the convergence and separation of swarm members.
\item Center of gravity: Aiming to minimize the distance from the central point to other points in the spatial distribution of the swarm.
\item Directional Accuracy: Measuring the accuracy between the swarm's movement and the desired travelling path.
\item Flock thickness: Measuring the swarm's density.
\item Resource depletion: Qualifying the irreversible consumption of limited resources by swarm members.
\item Swarm health: Evaluating the current status of the swarm.
\end{itemize}

In particular, swarm health is an important aspect for determining the difficulties faced by the swarm, and it can be decomposed following the analysis in~\cite{harriott2014biologically} into the following sub-components:

\begin{itemize}
\item Number of stragglers: studied in \cite{parrish2002self} as the number of fish of a school distant at least 5 body lengths from any other fish. This can reflect difficulties encountered by the swarm like obstacles in the environment or conflicting commands.

\item Subgroups number and size: as explained in \cite{navarro2009proposal}, the number and size of subgroups can vary due to obstacles or as a way to perform the task more efficiently. In a swarm, subgroups can be identified and measured using clustering algorithms.

\item Collision count: also from \cite{parrish2002self}, this is the number of collisions between members of the swarm. If a collision avoidance system is in place, this could be the number of times this system had to intervene.
\end{itemize}

Another factor that should be accounted for in an HSI system is \emph{Neglect Benevolence}, which is a consequence of the fact that a swarm needs some time to stabilize after receiving an instruction before being ready to receive further instructions. In \cite{nagavalli2014neglect} this concept is formally defined and analyzed, leading to a complex algorithm for finding the optimal intervention time that requires computing the convergence time for the swarm with input given at different times. In practice, it may be possible to estimate the current value of \emph{Neglect Benevolence} empirically utilizing the time since the last human intervention and the factors that \cite{walker2012investigating} reported to be influenced by \emph{Neglect Benevolence}: \emph{Directional Accuracy} and \emph{Cohesion}.

\begin{figure*}[tp]
\centering
  \includegraphics[width=0.80\textwidth]{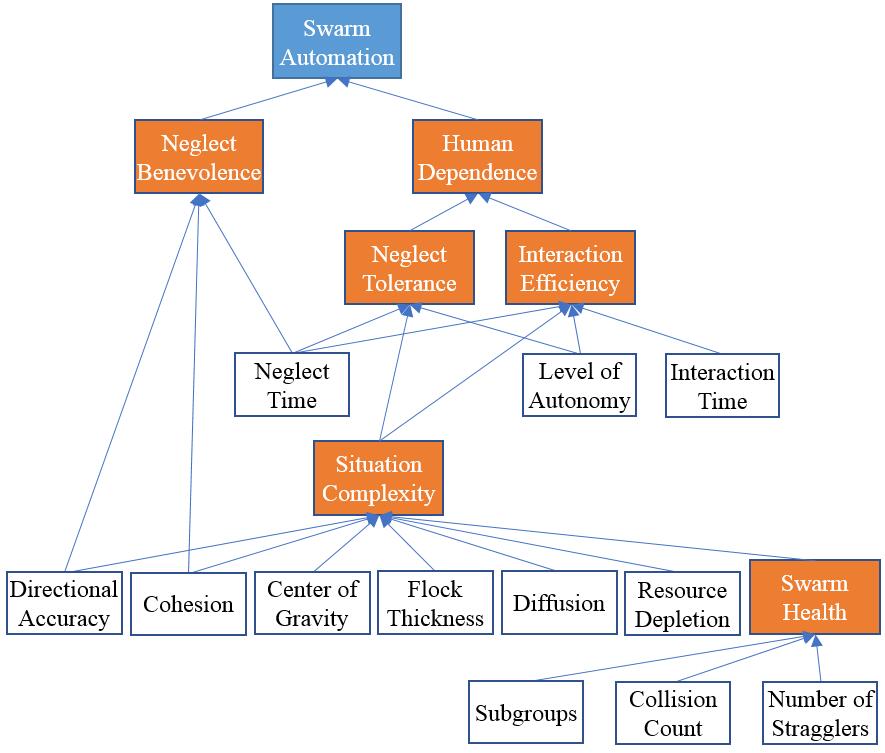}
  \caption{Metrics for Swarm Automation Level.}
  \label{fig:tree_automation}
\end{figure*}

The concepts introduced in this section are composed in Fig~\ref{fig:tree_automation}, showing how they can be combined to build an estimation of the \emph{Swarm Automation Level}.

\subsection{Human States}

An effective adaptive system for HST should be able to appropriately adjust its behaviour to fit the current situation, based on information from humans, swarm, and context. We will not consider possible physical interaction between a human and a swarm to be within the scope of this paper. Instead, we will limit our scope to problem solving tasks, where the physical interaction (through keyboard, mouse, joysticks, or the alike) imposes minimum load on the human that could be neglected, but where the cognitive load plays more substantial role on human performance.

Integrating human cognitive states into adaptive systems is a critical step towards effective and efficient human-swarm teaming for two reasons. First, real-time assessment of human's cognitive states, such as cognitive workload, fatigue and attention, enables the system to adjust itself to maintain the human states within a safe envelope. This is particularly useful in scenarios where human mistakes caused by overload/underload or fatigue could potentially result in hazardous consequences. Second, human cognitive states can be translated into meaningful guidance for adaptation (e.g., swarm level of autonomy). It becomes pertinent to the adaptive HST system to have a clear understanding of human cognitive states. While there is a myriad of studies that rely on subjective metrics for workload, they are unsuitable for real-time adaptation. For example, NASA-TLX (\cite{hart1988development}) is a very well-known subjective method to estimate workload. However, this method can not be used in real-time applications despite its common use in off-line analysis of experiments. 

In this section, we focus real-time assessment of human's cognitive states using psycho-physiological markers.
There exist several modalities for the objective measurement of human cognitive states. Electroencephalography (EEG) can be considered as one of the most common modality to estimate cognitive demands. In many studies frequency domain of the EEG signals has been used to estimate human's mental workload~\cite{borghiniameasuring}. Cognitive loads can affect power (e.g., power spectral density) of EEG spectrum in different frequency bands (e.g., theta). Event related potential (ERP) is also sensitive to the changes of mental workload. Heart rate (HR) is another form of physiological signal that can be used as a mental workload indicator. However, HR can be vulnerable  to detect other forms of workload as well (e.g., physiological activity). Electromyography (EMG) has been applied to analyze the mental workload of the human operator monitoring traffic density \cite{fallahi2016effects}. Transcranial Doppler sonography and functional near infrared (fNIR) are two hemodynamic methods that have been considered in mental workload analysis. In the Doppler sonography technique, cerebral blood flow velocity (CBFV) is correlated with the mental activity required by tasks. Eye-related measures such as pupillometry, fixations, and blinks have been applied to determine the driver mental workload. For instance, it has been shown that there is a positive relationship between the workload and blink latency (for review refer to \cite{marquart2015review}).

EEG is emerging is the type of modality that will likely offer a more robust and practical estimates of workload in a real-time environment. The temporal resolution of the EEG output is appropriate for real-time applications. Zhao et al.~\cite{zhao2012electroencephalogram} demonstrated the effect of mental fatigue on EEG signals recorded during a simulated driving task. They found that relative power (ratio between the power of each band and the power of total band) in theta band increased in the occipital, frontal and central regions as an effect of fatigue (Note that the relative power in each region was calculated by averaging relative powers of all electrodes in the corresponding region). In addition the relative power of alpha band increased in four regions namely parietal, temporal, central, and occipital. On the other hand, Beta rhythm saw a decrement in the temporal, frontal, and central regions. The work also investigated the effect of mental fatigue on event related potential (ERP). They reported a significant decrease in the amplitude of P300 in Fz and Cz electrodes.

EEG rhythms have been used to investigate the difference between strategies taken by expert and novice shooters \cite{doppelmayr2008frontal}. It was demonstrated that the power of theta band in frontal midline is related to focused attention. In the study conducted by Matthews et al. \cite{matthews2015psychometrics} workload indicators were extracted from subjects performing threat and change detection task scenarios. These two scenarios were also combined to analyze the workload metrics in dual task scenarios. Based on their results, EEG Task Load Index (TLI, ratio of theta power at Fz to alpha power at Pz) metric can be used to distinguish between the dual-task and single-task performances. In another study, the entropy of discrete wavelet transform (DWT) coefficients extracted from EEG signals was used to estimate seven levels of the cognitive workload (\cite{zarjam2013estimating}). However, this method of mental workload estimation could not be appropriate for real-time applications and as a result is not considered in our proposed framework.

Moreover, EEG and ERP indicators were used to switch between two modes of performing tasks, manual and automated, as well as to explore effectiveness of adaptive autonomy in the task performance (\cite{prinzel2003effects}). EEG engagement index ($20 \times \textit{beta}/(\textit{alpha}+\textit{theta})$) was used to estimate the mental workload of participants performing multi attribute battery tracking task, and in turn to determine the task mode. Participants involved in the adaptive autonomy group showed better performance and lower level of subjective workload scores compared to the control groups, which indicates the efficacy of this EEG index for the estimation of mental workload. In the same study P300 was also analyzed when participants were asked to perform a second task (auditory oddball task) while performing the first task. It has been shown that the amplitude of P300 is sensitive to the mental resources available to perform a task (\cite{isreal1980p300}). Therefore, P300 was used as another indicator of effectiveness of the adaptive autonomy. The adaptive autonomy group had a better performance in auditory oddball task compared to control groups and showed a larger P300 amplitude. This results prove that performing task in adaptive autonomy mode can free more attentional resources for performing a secondary task. In our proposed framework these two mental workload metrics, namely  $20 \times \textit{beta}/(\textit{alpha}+\textit{theta})$ and P300, can be used to change the level of autonomy in multitask scenarios. The former index, along with other related indicators, will be used to adapt the interaction related to the main task (ongoing task), while the latter one can be used to adapt the interaction related to secondary tasks (interrupting tasks).

\begin{figure*}[tp]
\centering
\includegraphics[width=0.80\textwidth]{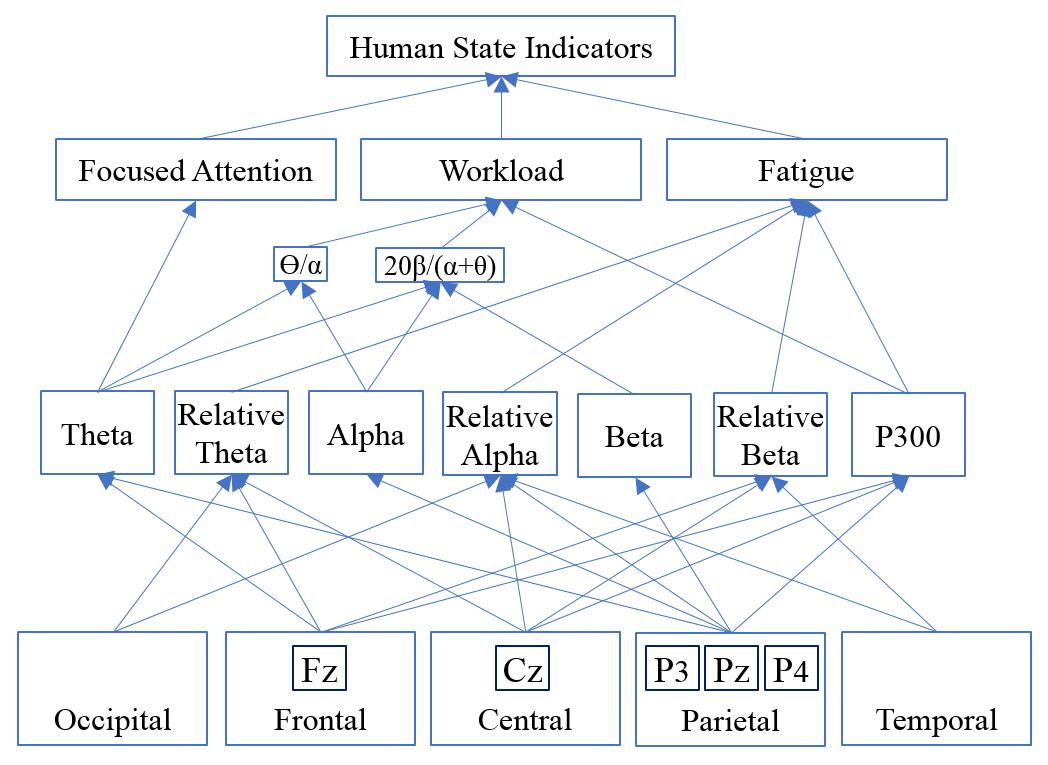}
\caption{The human cognitive state indicators.}
\label{fig:tree_human}
\end{figure*}

Figure \ref{fig:tree_human} summarizes the relevant brain regions and sensor locations on the scalp and significant frequency bands and peaks to measure three human cognitive state indicators, namely focused attention, workload and fatigue from the existing literature.

\section{Synthesis: The MICAH Framework}
\begin{figure}[tp]
\centering
\includegraphics[width=0.48\textwidth]{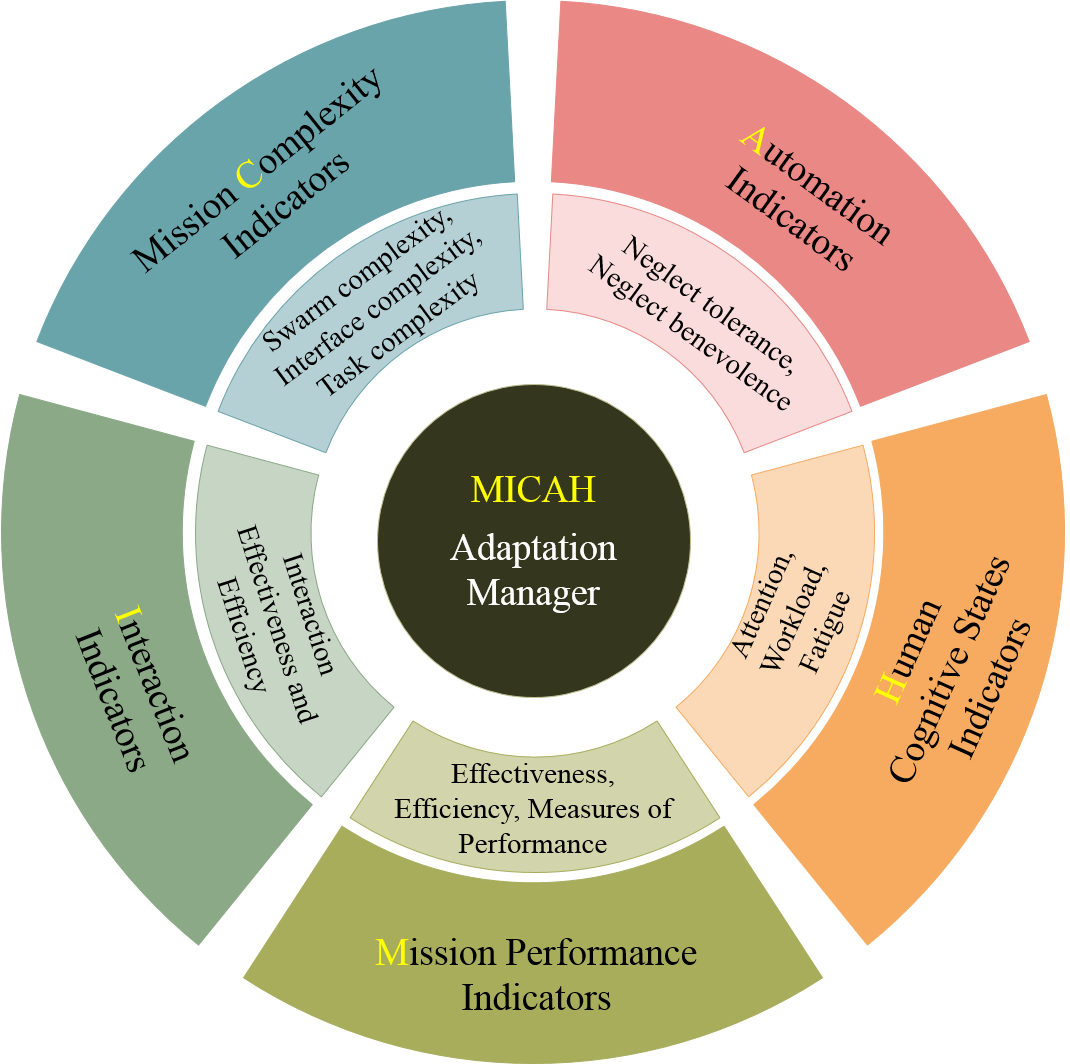}
\caption{MICAH: categories of indicators used for adaptation in HST}
\label{fig:MICAH}
\end{figure}

In the previous sections we described five different types of indicators, showed how they can be computed with methods already known in the literature, and highlighted why they should be relevant for an adaptive HST system. We believe that an effective adaptive HST system should contain five components, each addressing one of the five types of indicators; we named this structure with the acronym \emph{MICAH}, and a visual summary of it is presented in Figure \ref{fig:MICAH}.
The five types of indicators can be summarized in the following way with the letter contributing to the abbreviation MICAH underlined:

\begin{itemize}
  \item \underline{M}ission Performance: composed of effectiveness and efficiency, it is the primary objective of the system and should never be disregarded;

  \item \underline{I}nteraction: describes how productive the interaction between the human and the swarm is; monitoring this indicator gives insight into the current interaction mode;

  \item Mission \underline{C}omplexity: studies how the task, interface and swarm contribute to the workload for the human; it is an important factor in determining the performance of the human and the complexity of the mission at a particular point of time to trigger appropriate level of automation;

  \item \underline{A}utomation level: analyzes the performance of the swarm and its need for human intervention, which are fundamental inputs to correctly set the level of autonomy;

  \item \underline{H}uman cognitive states: assesses the mental conditions of the human, determining for example if they are overloaded or underloaded and allowing the system to adapt accordingly
\end{itemize}

The main purpose of MICAH is being a synthesis of indicators needed to design adaptive Human-Swarm Teaming systems. A practical system does not need to use the exact indicators described in this paper, but the five components should all be considered by the adaptation manager.

An example of adaptive HST could be a setup where a swarm needs to patrol an area containing a number of checkpoints, and to succeed each checkpoint needs to be visited repeatedly with a delay no higher than a fixed threshold. The human interacts with the swarm by teleoperating one or more robots, and the autonomy of the swarm is determined by the number of robots under direct control of the human. In this case, the adaptive manager could consider the following indicators:
\begin{itemize}
\item Mission Performance: Number of checkpoints reached in the time limit, average time to revisit a checkpoint
\item Interaction: Performance of robots guided by the human compared to the autonomous ones
\item Mission Complexity: Number of subswarms and current level of autonomy
\item Automation: Alignment and cohesion of the swarm
\item Human cognitive states: Workload perceived by the human
\end{itemize}

By assessing these indicators all the components of MICAH are considered, and the system has all the information needed to correctly deliberate on the best adaptation.


The closed-loop adaptive systems this paper targets are the ones that can appropriately modify their behaviors to fit the needs of the current context, to meet the changing needs of the human operators, while maintaining good performance, without explicit instructions from the human operators. The purpose of designing such adaptive systems for human-swarm interaction is to combine the capabilities of human and swarms to maximize their potentials, in order to achieve effective and efficient teamwork.

\begin{figure*}[ht!]
\centering
\includegraphics[width=0.95\textwidth,height=0.5\textheight]{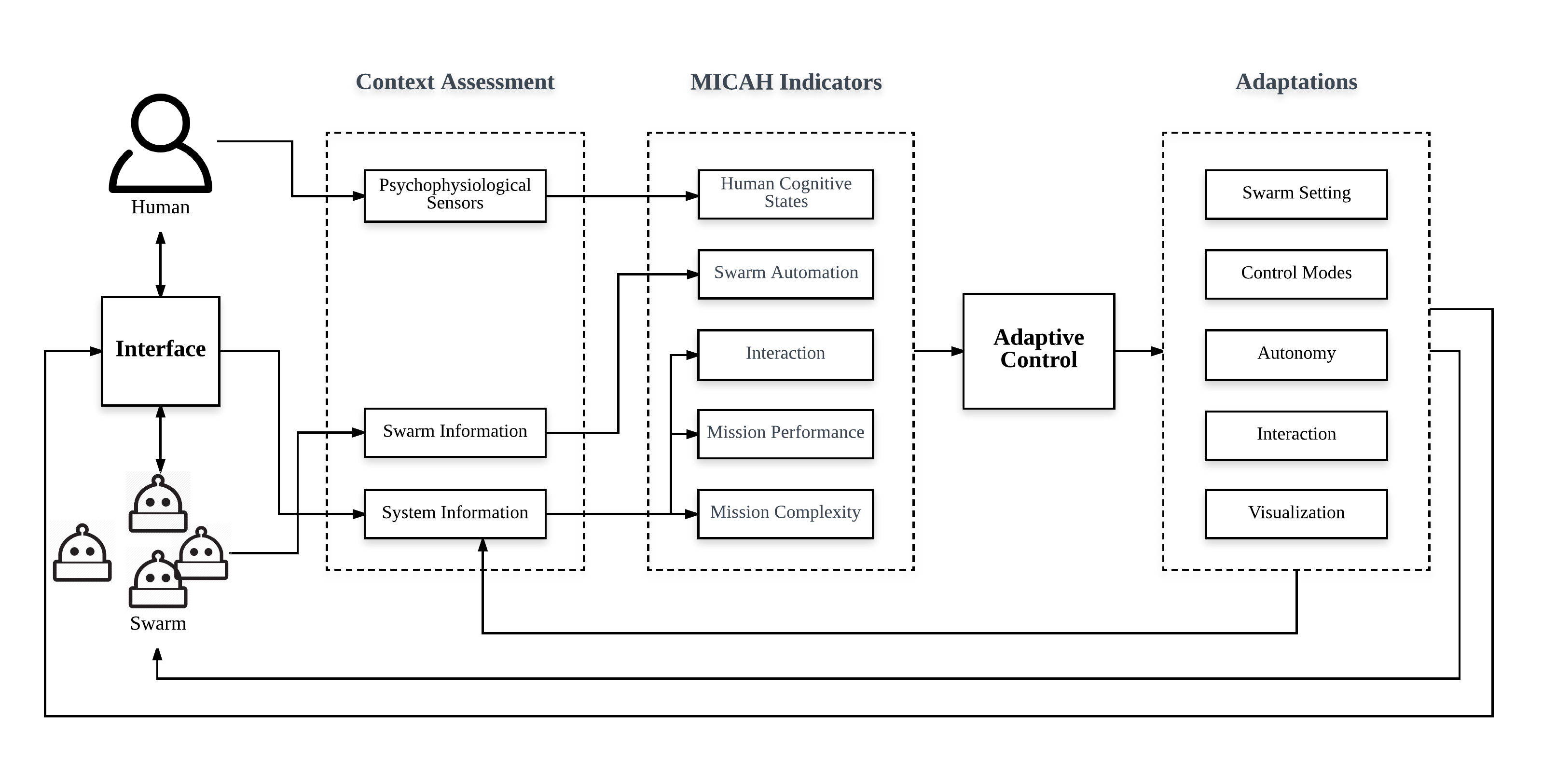}
\caption{Diagram of the closed-loop adaptive system for human-swarm teaming.}
\label{fig:framework}
\end{figure*}

To illustrate the framework of the adaptive system for HST, Figure~\ref{fig:framework} depicts the system flow diagram, taking a pattern of four steps: context assessment, MICAH indicators, adaptive control and adaptation. With the four-step cycle, the framework is able to sense the context information, including the human's cognitive states, the swarm states, the mission, system and environment states; integrate and translate these information into representative indicators, the MICAH indicators; update adaptive control dynamically based on the indicators; then finally adopt appropriate adaptations in five aspects, namely swarm setting, control modes, autonomy level, interaction modes and visualization. 

The dynamic updates of these aspects enables the system to adjust itself to meet the human's cognitive requirements, while ensuring mission success. For example, the psychophysiological sensors in the context assessment module collects the human operator's cognitive information, which is translated into a series of human states indicators, including workload indicator, fatigue indicator and focused attention indicator. Integrated with information of the current task and system states, these human states indicators (workload, fatigue and attention) are used by the adaptive controller to decide adaptations. For example, high workload and fatigue might compromise the human operator's performance so the system raises the autonomy level of swarm to let the human operator only focus on the most critical task. While, lack of attention might cause serious consequence in the case of emergencies therefore the system lowers autonomy level and updates interaction modes and visualization to counterbalance the human states within safe range.

Conventional adaptive systems trigger adaptations in specific situations or for particular tasks, return to regular operation and disengage the activation once the situation or task is finished. However, as the situation and context evolve, the once adequate adaptation strategy applied by conventional adaptive systems might become inadequate~\cite{fuchs2017towards}. To address this problem, the MICAH framework provides dynamic adaptations based on indicators from extensive aspects, therefore is capable of adjusting itself to the situation, the human operator and the context. The MICAH framework acts as an intelligent assistant that unobtrusively observes and comprehends the human operator's actions, cognitive states and the evolving context and situations, and provides appropriate adjustments automatically. The adaptations are flexible and dynamic and can counterbalance the human states and maximize the overall performance.

\section{Conclusion}

This paper leverages literature from human machine interaction, swarm intelligence and adaptive systems, focusing on identifying significant information for designing closed-loop adaptive systems for effective and efficient bi-direction communication in human-swarm teaming. It defines the core concepts of HST and proposes an integrated model, the MICAH framework, for bringing together the multitude of indicators required to design closed-loop adaptive systems. We began with a discussion defining human-swarm teaming and its properties, then identified some major challenges in HST with particular attention to autonomy and closed-loop adaptive systems. Five groups of indicators, summarized as MICAH, were proposed for adaptive control to monitor and balance human mental states while maximizing the human-swarm teaming potential to complete the mission successfully. 

The MICAH framework extends existing concepts of adaptive systems, which mainly focused on task allocation, to fit swarm systems in order to achieve effective and efficient human-swarm teaming. The main features of this line of research are summarized as follows.

\begin{enumerate}
\item We take the evolving context, human's changing cognitive states and overall performance into consideration for adaptive control, with a particular focus on human-swarm teaming. Mission performance factors, interaction factors, complexity factors, swarm-related factors, and human-related factors are for the first time integrated together for adaptation decisions.

\item We identified five groups of indicators, summarized as MICAH, for HST. MICAH is capable of sensing the context information, including human's cognitive states, swarm states, mission states, and system states; integrating and translating this information
into representative indicators for adaptation in swarm setting, control modes, autonomy level, interaction modes, and visualization.

\item  From the perspective of autonomy, MICAH supports dynamic and flexible adaptations that can be used in both adaptive autonomy systems and mixed-initiative systems. From the perspective of adaptive control, MICAH supports seamless supervision or monitoring of the human states and system that it can adapt itself to fit the evolving context, human's changing cognitive states and overall performance. This flexible adaptive control is an extension to current once adequate adaptations.
\end{enumerate}

%
%
%
%
%


\end{document}